\documentclass[preprint2,tigten,letteredappendix,appendixfloats]{aastex7}

\usepackage{amsmath}
\usepackage{romannum}
\usepackage{natbib}
\usepackage{enumitem}
\usepackage{soul}
\usepackage{longtable}

\usepackage{graphicx}           
\usepackage{latexsym}           
\usepackage{bm}                 
\usepackage[english]{babel}
\usepackage{color}              

\def\arrvline{\hfil\kern\arraycolsep\vline\kern-\arraycolsep\hfilneg}

\begin{document}

\pagenumbering{arabic}


\title{Point-like and Narrow-lined: A Potentially New Population of Objects Discovered by JWST}

\author[0000-0001-7592-7714]{Haojing Yan}
\affiliation{Department of Physics and Astronomy, University of Missouri, Columbia, MO 65211, USA}
\email{yanha@missouri.edu}

\author[0000-0001-7957-6202]{Bangzheng Sun}
\affiliation{Department of Physics and Astronomy, University of Missouri, Columbia, MO 65211, USA}
\email{bangzheng.sun@mail.missouri.edu}

\author[]{Riley Shive}
\affiliation{Department of Physics and Astronomy, University of Missouri, Columbia, MO 65211, USA}
\email{rjshvd@missouri.edu}

\begin{abstract}

    We report a potentially new population of objects revealed by the data from
the James Webb Space Telescope, which are characterized by their point-like 
morphology and narrow permitted emission lines. Our sample includes eight 
objects found in three JWST wide survey fields, which have $z=3.624$ to 5.378 
and $M_B \approx -18.3$ to $-20.7$~mag. Their light distributions follow 
Gaussian profiles, with the full-width-at-half-maximum (FWHM) values only 
3.7\%--35.6\% larger than those of the point spread functions. Their sharpest 
FWHM sizes as measured in the bluest bands correspond to only 0.49 to 0.96~kpc. 
They have very strong [O~\Romannum{3}] and H$\alpha$ lines (median rest-frame 
equivalent widths of 1804 and 1460~\AA, respectively), and the line widths of 
the latter are only 150--360~km~s$^{-1}$. Due to the limitation of the current 
data, the 
exact nature of this new population is still uncertain. The emission line 
diagnostics show that at least one object is consistent with being AGN. On 
the other hand, the spectral energy distributions of most of the eight objects 
can be fitted reasonably by normal galaxy templates, which suggest that they 
could also be very young (median age of 120~Myrs), star-forming (median star 
formation rate of  1.7~$M_\odot$~yr$^{-1}$) galaxies in the early formation 
stage (having acquired a median stellar mass of only $10^{8.4}M_\odot$). If 
they are indeed star-forming galaxies, their gas-phase metallicities range 
from 12+log(O/H)~$= 8.1$ to 8.3. It will be critical to understand this 
population by deeper medium-resolution spectroscopy in the future. If they are 
AGNs, they constitute a new kind of type~2 AGNs that are low-luminosity and 
almost ``hostless''. If they are star-forming galaxies, they also constitute a 
new kind whose early formation is likely a secular process starting from a 
very compact core.

\end{abstract}


\section{Introduction}

   Generally speaking, point sources seen in imaging surveys are mostly 
Galactic stars, and a small fraction of them are quasars. Quasars being
point sources is only because the strong emission from their accreting disks,
which cannot be resolved by the current instruments, swamps the light from 
their host galaxies. 
   
   Our study presented here was motivated by the recent discovery of a new
population of extragalactic objects dubbed as ``little red dots'' (LRDs), 
which was made by the James Webb Space Telescope (JWST). LRDs are very 
compact, have ``V-shaped'' spectral energy distributions (SEDs) and have 
permitted lines with widths of $\sim$1000--2000~km~s$^{-1}$, which are at the 
boarder line of type~1 AGNs 
\citep[see e.g.,][]{Labbe2023, Kocevski2023, Harikane2023b, Barro2024, Williams2024, PerezGonzalez2024b, Matthee2024, Greene2024}.
The true nature of LRDs, especially their broad-line mechanism, is still under 
debate, and it is not our purpose to study them in this paper. It was the 
fact that a significant fraction of LRDs being point-like in the 
high-resolution JWST images inspired a simple question that we intended to
address: are there any other new kinds of objects among point-like sources 
that we did not notice previously?

   To this end, we selected a large sample of $\sim$2000 point-like sources
in three wide JWST survey fields and cross-matched them with the objects that we 
identified using the JWST spectroscopic data from various programs. The study of 
the full sample will be deferred to a future paper. In this work, we present a 
subsample of eight objects that represent a potentially new population of 
extragalactic sources, which are characterized by their point-like morphology 
and narrow permitted emission lines with widths of $<400$~km~s$^{-1}$. The paper 
is organized as follows. The data used for this study are presented in 
Section~2. The details of the source morphology and the emission line 
measurements are given in Section~3. We discuss the possible nature of these 
objects in Section~4 and summarize in Section~5. All magnitudes quoted are in 
the AB system, and all coordinates are in the ICRS frame and Equinox 2000. We 
adopted a flat $\Lambda$CDM cosmology with $\rm H_0=71~km~s^{-1}~Mpc^{-1}$, 
$\rm \Omega_{M}=0.27$, and $\rm \Omega_{\Lambda}=0.73$. 
   
\section{Data Description}

\subsection{Imaging Data}

   We used the NIRCam data from two JWST Cycle 1 wide-field survey programs, 
namely, the ``Cosmic Evolution Early Release Science Survey'' 
\citep[CEERS; PID 1345; PI S. Finkelstein][]{Finkelstein2023} in the EGS 
field, and the ``Public Release IMaging for Extragalactic Research'' (PRIMER) 
program (PID 1837; PI J. Dunlop) in the UDS and the COSMOS fields. The CEERS 
program used seven NIRCam bands: F115W, F150W, and F200W in the
short-wavelength (SW) channel, and F277W, F356W, F410M, and F444W in the 
long-wavelength (LW) channel. The coverage is 86.45~arcmin$^2$. 
The PRIMER NIRCam observations utilized eight passbands: F090W, F115W, F150W, 
and F200W in the SW channel, and F277W, F356W, F410M, and F444W in the 
LW channel. The coverage in the UDS and the COSMOS fields are 186.8 and 
137.1~arcmin$^2$, respectively. 

    The reduction of these NIRCam images has been described in 
\citet[][]{Yan2023a} and \citet[][]{SY2025a}. For photometry, we created the 
mosaics with the pixel scale of 60~mas~pix$^{-1}$ (hereafter the ``60mas'' 
images). For the morphological analysis, we created small stacks with finer 
pixel scales around the sources of interest, which will be detailed in 
Section~3.1. 

    For the purpose of SED analysis, we also used the HST ACS F606W and F814W
60mas images from the CANDELS survey \citep[][] {Grogin2011,Koekemoer2011}. Our
NIRCam images are aligned to these ACS images.

    Following the methodology as described in \citet[][]{SY2025a}, the 
photometry was done by running {\sc SExtractor} \citep[][]{Bertin1996}
in the dual-image mode using the F356W images as the detection images.

\subsection{JWST Spectroscopic Data}

   Our spectroscopic data are mainly from two JWST NIRSpec multi-object 
spectroscopy programs utilizing the micro-shutter assembly (MSA). One is the ``Red Unknowns: Bright Infrared Extragalactic Survey'' 
\citep[RUBIES; PID 4233, PIs A. de Graaff \& G. Brammer;][]{RUBIES2025} in 
Cycle 2, and the other is the ``CANDELS-Area Prism Epoch of 
Reionization Survey'' (CAPERS; PID 6368; PI. M. Dickinson) in Cycle 3. The
observations of both programs used three-shutter slitlets and a three-point 
nodding pattern. The RUBIES program observed the same targets in both the 
low-resolution PRISM/CLEAR (hereafter ``PRISM'') and the medium-resolution 
G395M/F290LP (hereafter ``Grating'') disperser/filter setups with the same 
slit orientation. Under these setups, the resolving powers are 
$R\approx 30$--300 and $R \approx 1000$, respectively, and the wavelength 
coverages are 0.6--5.3~$\mu$m and 2.87--5.10~$\mu$m, respectively. The CAPERS 
program only used the PRISM setup. The reduction of these data has been 
described in \citet[][]{YSL2024}, \citet[][]{SY2025b} and \citet[][]{SY2025c}.
The final spectra extraction was done in the last step of the reduction using
the {\sc msaexp} package \citep[version 0.9.2;][]{Brammer_msaexp2023}.

   In addition, we also used the data taken by the NIRCam instrument in its 
wide-field slitless spectroscopy (WFSS) mode, which were from the  
``COSMOS-3D'' program (PID 5893; PI K. Kakiichi) in Cycle 3. These observations
used Grism-R in the F444W band, which resulted in medium-resolution spectra of
$R\approx 1600$ over 3.881--4.982~$\mu$m. To reduce these data, 
we retrieved the Level 1b data from the Mikulski Archive for Space Telescopes 
(MAST) and ran them through the {\tt calwebb\_detector1} step of the JWST data 
reduction pipeline 
\citep[][version 1.18.0 in the context of jwst\_1364.pmap]{Bushouse24_jwppl} 
and obtained the ``rate.fits'' files. We then followed the procedures of
\citet[][]{Sun2023} to further process these files and extract the spectra.
    
\section{Point-like, Narrow-line Sources and Characterization}

\begin{figure*}
    \centering
 \includegraphics[width=0.95\textwidth,height=0.9\textheight,keepaspectratio]{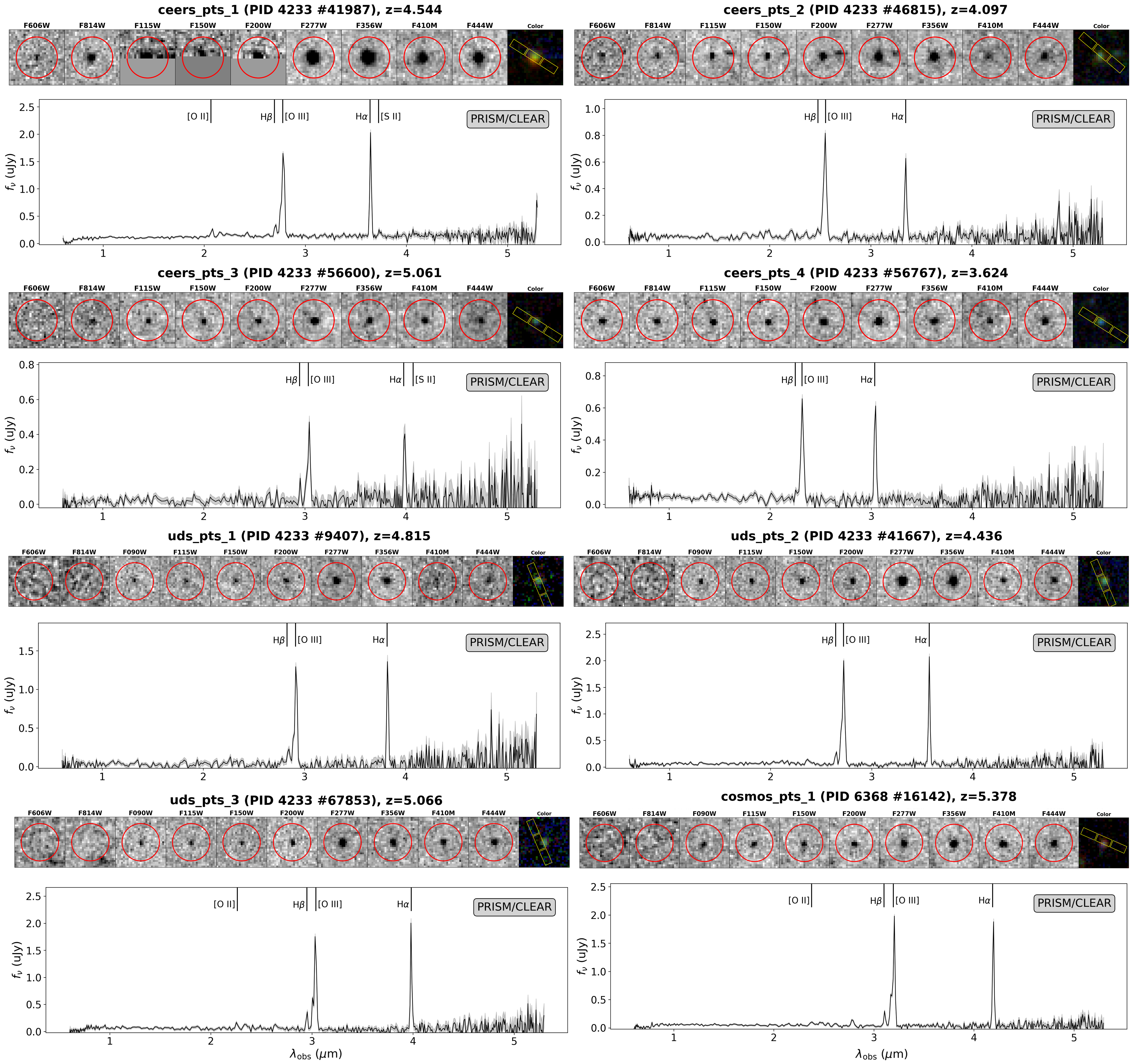}
\raggedright
    \caption{Image stamps and 1D spectra of the eight objects in our sample,
    with the object ID and the redshift labeled.
    The images are 1\farcs6 on a side, and the red circle is 0\farcs64 in radius. The
    passbands are as labeled. The F606W and F814W are from the HST ACS, while
    the rest are from the JWST NIRCam. The NIRSpec slit placement is overlaid 
    on the color image. The 1D spectra are the low-resolution PRISM spectra
    from either PID 4233 (for the ones in CEERS and UDS) or 6368 (for the one
    in COSMOS), and the target ID in the respective program is in the parenthesis.
    The detected lines are marked.
    }
    \label{fig:imgspec}
\end{figure*}

    The point-like sources were selected using a combination of the
\texttt{MAG\_AUTO} versus \texttt{FLUX\_RADIUS} diagnostics and the 
\texttt{CLASS\_STAR} parameter, all provided in the output from the photometry
of {\sc SExtractor}. The details will be given in a future paper. For this 
work, we matched the point-like sources to the objects identified in the
spectroscopic data and selected those that have secure H$\alpha$ detections. 
The H$\alpha$ line widths were measured using the medium-resolution spectra, 
and we retained only those that have full-width-at-half-maximum (FWHM) of 
$\Delta v < 1000$~km~s$^{-1}$. In the end, eight point-like, narrow-line 
sources made into our final sample, which span the redshift range of 
$z=3.624$--5.061. These sources are listed in Table~\ref{tbl:phot}, along with 
their ACS and NIRCam photometry. The {\sc SExtractor} \texttt{MAG\_ISO} 
magnitudes are adopted, and we use the notation of ``$m_{\rm wav}$'' to denote
the magnitudes in a given band, where ``wav'' is the three numbers in the band
designation indicating the wavelength; e.g., $m_{200}$ is the magnitude in
the NIRCam F200W. Figure~\ref{fig:imgspec} shows their image stamps and 
their PRISM spectra indicating the identifications. The analysis of their 
morphologies and line widths is detailed below.

\begin{table*}[hbt!]
    \centering 
    \small
    \caption{List of objects in the sample and their photometry }
    \resizebox{0.99\textwidth}{!}{
    \begin{tabular}{lcccccccc} \hline\hline
    ID & ceers\_pts\_1 & ceers\_pts\_2 & ceers\_pts\_3 & ceers\_pts\_4 & uds\_pts\_1 & uds\_pts\_2 & uds\_pts\_3 & cosmos\_pts\_1 \\
    \hline
    Spec ID & 4233\_41987 & 4233\_46815 & 4233\_56600 & 4233\_56767 & 4233\_9407 & 4233\_41667 &  4233\_67853 & 6368\_16142 \\
    \hline 
    $z_{\rm spec}$ & $4.544\pm0.001$ & $4.097\pm0.001$ & $5.061\pm0.001$ & $3.624\pm0.001$ & $4.815\pm0.001$ & $4.436\pm0.001$ & $5.066\pm0.001$ & $5.378\pm0.001$ \\
    \hline
    R.A.  & 214.8776215 & 214.9616869 & 214.9748850 & 214.8968405 & 34.2686220 & 34.2917549 & 34.2739105 & 150.1315787 \\
    Decl. & 52.8481446 & 52.9214407 & 52.9526796 & 52.8969789 & $-$5.3001699 & $-$5.2437932 & $-$5.1938808 & 2.3903229 \\
    \hline 
    $m_{606}$ & $27.90\pm0.29$ & $28.44\pm0.36$ & $>29.19$ & $27.57\pm0.08$ & ... & $28.00\pm0.26$ & $>28.84$ & $>27.92$ \\
    $m_{814}$ & $26.68\pm0.09$ & $27.69\pm0.19$ & $28.76\pm0.21$ & $27.64\pm0.11$ & ... & $27.06\pm0.12$ & $27.45\pm0.12$ & $28.19\pm0.85$ \\
    $m_{090}$ & ... & ... & ... & ... & $27.27\pm0.27$ & $27.56\pm0.25$ & $27.17\pm0.24$ & $27.75\pm0.22$ \\
    $m_{115}$ & ... & $27.62\pm0.12$ & $28.25\pm0.12$ & $27.51\pm0.09$ & $27.60\pm0.38$ & $27.29\pm0.19$ & $27.25\pm0.25$ & $27.45\pm0.17$ \\
    $m_{150}$ & ... & $27.72\pm0.17$ & $28.26\pm0.14$ & $27.66\pm0.12$ & $27.21\pm0.23$ & $26.84\pm0.10$ & $27.67\pm0.32$ & $27.70\pm0.18$ \\
    $m_{200}$ & ... & $27.23\pm0.07$ & $28.47\pm0.09$ & $27.65\pm0.05$ & $27.16\pm0.18$ & $27.08\pm0.11$ & $27.22\pm0.16$ & $27.70\pm0.15$ \\
    $m_{277}$ & $25.56\pm0.02$ & $26.69\pm0.05$ & $27.76\pm0.08$ & $27.50\pm0.08$ & $26.57\pm0.06$ & $26.25\pm0.03$ & $26.69\pm0.05$ & $27.54\pm0.08$ \\
    $m_{356}$ & $25.68\pm0.02$ & $26.98\pm0.05$ & $28.93\pm0.20$ & $28.26\pm0.14$ & $27.13\pm0.09$ & $26.75\pm0.04$ & $27.51\pm0.10$ & $27.09\pm0.05$ \\
    $m_{410}$ & $25.94\pm0.06$ & $27.40\pm0.17$ & $28.02\pm0.17$ & $28.35\pm0.29$ & $27.87\pm0.35$ & $27.89\pm0.26$ & $26.99\pm0.14$ & $26.98\pm0.09$ \\
    $m_{444}$ & $26.02\pm0.04$ & $27.49\pm0.11$ & $28.59\pm0.13$ & $28.29\pm0.09$ & $27.85\pm0.24$ & $27.57\pm0.13$ & $27.27\pm0.13$ & $27.55\pm0.10$ \\
    \hline
    $M_B$ & $-20.7$ & $-19.2$ & $-18.7$ & $-18.3$ & $-19.8$ & $-20.0$ & $-19.8$ & $-19.0$ \\
    \hline 
    \end{tabular}
    }
\raggedright
    \tablecomments{``Spec ID'' is made of the NIRSpec program ID (4233 or 
    6368) followed by the target ID in the respective program. ``$M_B$'' is
    the absolute magnitude in the rest frame B-band; depending on the source
    redshift, it is calculated using either $m_{200}$ or the interpolation of 
    $m_{150}$ and $m_{200}$.
    }
    \label{tbl:phot}
\end{table*}

\subsection{Point-like morphology}\label{sec:morph}

    To verify their point-like morphology, we compare their sizes to those of
the point spread functions (PSFs). For this purpose, we first constructed the 
empirical PSFs (hereafter ``EPSFs'') in the relevant bands in each field. The
procedure and the results are given in Appendix~\ref{sec:epsf}. The pixel
scales of these EPSF images are 20~mas~pix$^{-1}$ and 30~mas~pix$^{-1}$ in 
the SW and LW bands, respectively.
    
\begin{figure*}
    \centering
    \includegraphics[width=0.95\textwidth]{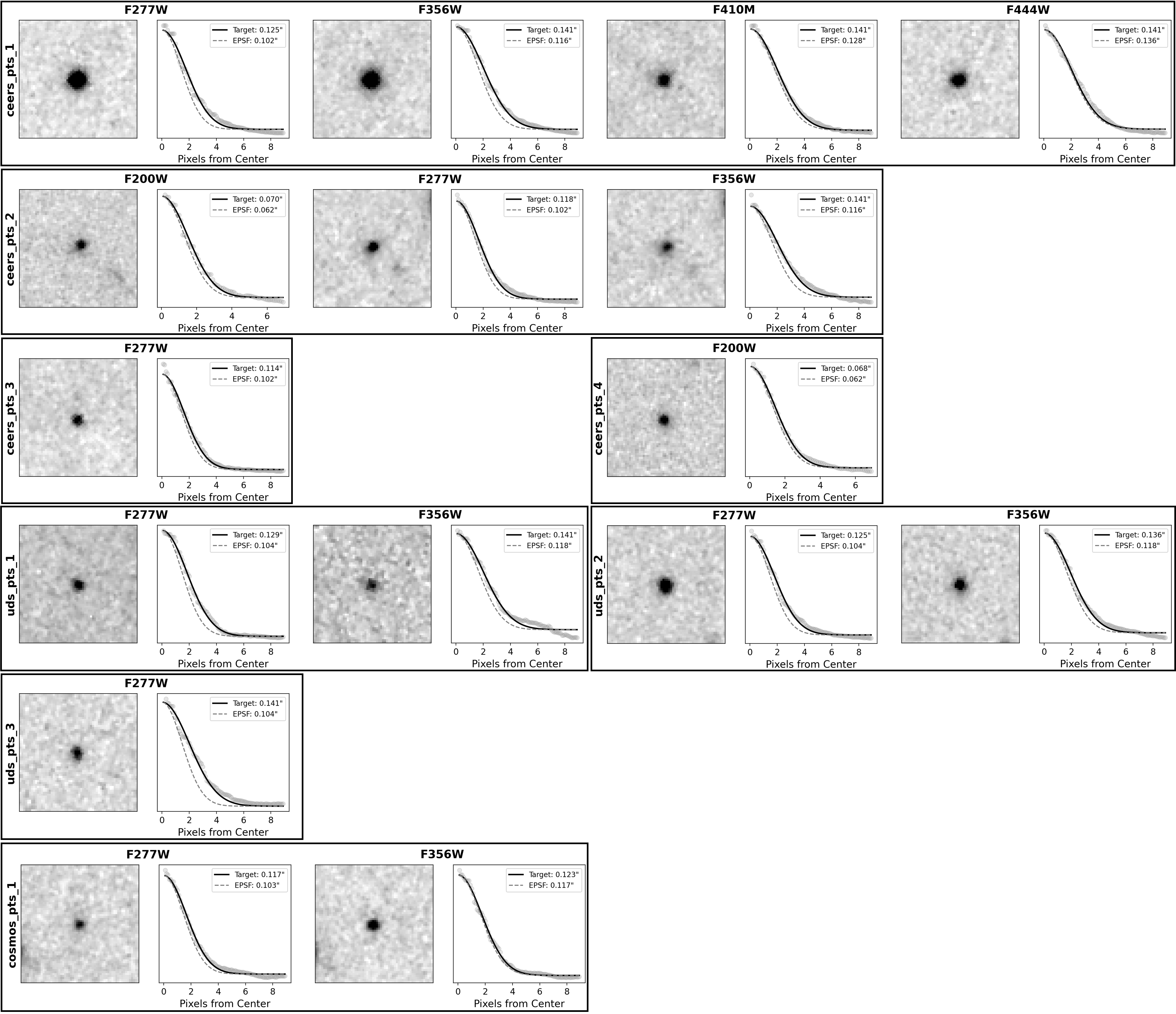}
    \raggedright
    \caption{Point-like morphologies of our objects. For clarity, the figures
    pertaining to a given object are framed by black border lines. For each
    object, we only show its image(s) in the band(s) where it has 
    ${\rm S/N\geq 10}$. The images have the pixel scale of either 20~mas
    (in the SW channel) or 30~mas (in the LW channel). The plot next to an
    image shows the best-fit Gaussian profile (solid black curve) to the light
    distribution (gray solid circles), together with the Gaussian profile that
    represents the EPSF (dashed gray curve) as described in 
    Appendix~\ref{sec:epsf}. The FWHM values of the best-fit Gaussian
    profiles and those of the EPSFs are labeled.
    }
    \label{fig:2dgauss}
\end{figure*}

    We then generated new images around our targets, adopting the same pixel 
scales as the EPSF images. These small images are centered on the targets to 
minimize the possible distortion due to the projection effect. We fitted a 2D 
Gaussian profile to their light distributions in the same way as we did for 
the EPSFs. As such a fit is not reliable on low S/N images, this was only 
done in the bands where the targets have ${\rm S/N}\geq 10$ as measured in a 
Kron aperture (corresponding to \texttt{MAG\_AUTO} measured by 
{\sc SExtractor}). Figure~\ref{fig:2dgauss} shows these results and compares 
to the fits of the EPSFs.

    As evident from this figure, our targets can be well described by a 2D Gaussian 
function. Their light profiles are rather close to those of the EPSFs, with the FWHM 
values being only 3.7\%--34.6\% larger than those of the respective EPSFs. Ideally, 
such a comparison should be done in the SW bands because they have sharper 
PSFs than the LW ones (in particular in F115W where the JWST optimizes the 
PSF size). Two sources, \texttt{ceers\_pts\_2} and \texttt{ceers\_pts\_4},
have sufficiently high S/N in one SW band, F200W. Their FWHM sizes are only
12.9\% and 9.7\% larger than the EPSF, respectively. 

    In short, while these eight objects are not exactly point sources, they 
are very close to being point-like (especially \texttt{ceers\_pts\_2} and 
\texttt{ceers\_pts\_4}). Based on their smallest FWHM values, the physical 
FWHM sizes of these eight objects correspond to 0.49--0.96~kpc. For a
Gaussian profile, its FWHM is exactly twice the effective radius ($R_e$); in
other words, their $R_e$ range from 0.25 to 0.48~kpc.

\subsection{Line measurements and Narrow H$\alpha$ widths }
\label{sec:nline}

    We measured the line properties of our targets following the methodology 
of \citet[][]{SY2025b}. Briefly, we fitted the continuum of a spectrum with a 
seventh-order Chebyshev polynomial using the {\tt fit\_generic\_continuum} 
utility in the {\sc astropy/specutils} package \citep[][]{Earl24_specutils}. 
When performing the fit, we excluded the wavelength regions that the emission 
lines fall into. The fitted continuum was then subtracted from the spectrum for
line measurement. 

    For the sake of consistency, the line intensities were measured on the 
low-resolution PRISM spectra, because they cover all the lines. The measurement
was mostly done by fitting a Gaussian profile to each emission line and 
integrating the best-fit profile within $2\times{\rm FWHM}$ around the fitted 
central wavelength. In a few cases, the lines fall in the lowest resolution 
part of the PRISM spectra ($\sim$1--2~$\mu$m) and/or are weak, 
and there are no sufficient 
resolution elements to perform the Gaussian fit. For such lines, their 
intensities were obtained by summing the values of the pixels occupied by the 
lines (only one or two pixels along the dispersion direction). The line 
equivalent widths (EW) are also measured. Table~\ref{tbl:lines} 
summarizes the results. If a line is not detected, the 2~$\sigma$ noise level 
at the line location is quoted. Note that the [S~\Romannum{2}] doublets are 
not resolved in the PRISM spectra, and the lines were treated as single lines. 
The situation for the [O~\Romannum{3}] doublets is more complicated. In some
cases they are partially resolved, and we fitted two Gaussian profiles
simultaneously to deblend. In other cases they are unresolved, and we treated
them as single lines; the total values are reported under 
[O~\Romannum{3}]$\lambda$5007 in the table. As the MSA slits did not 
fully cover the targets, the measured intensities suffer from the slit loss. 
We derived the correction for this loss based on the slit coverage, and the 
procedure is detailed in Appendix~\ref{sec:slitloss}. The correction factors 
are given in the last column of Table~\ref{tbl:lines}; the quoted 
intensity and luminosity values should be multiplied by these factors to 
recover the full line intensities and luminosities.

\begin{figure*}[hbt!]
    \centering
    \includegraphics[width=0.95\linewidth]{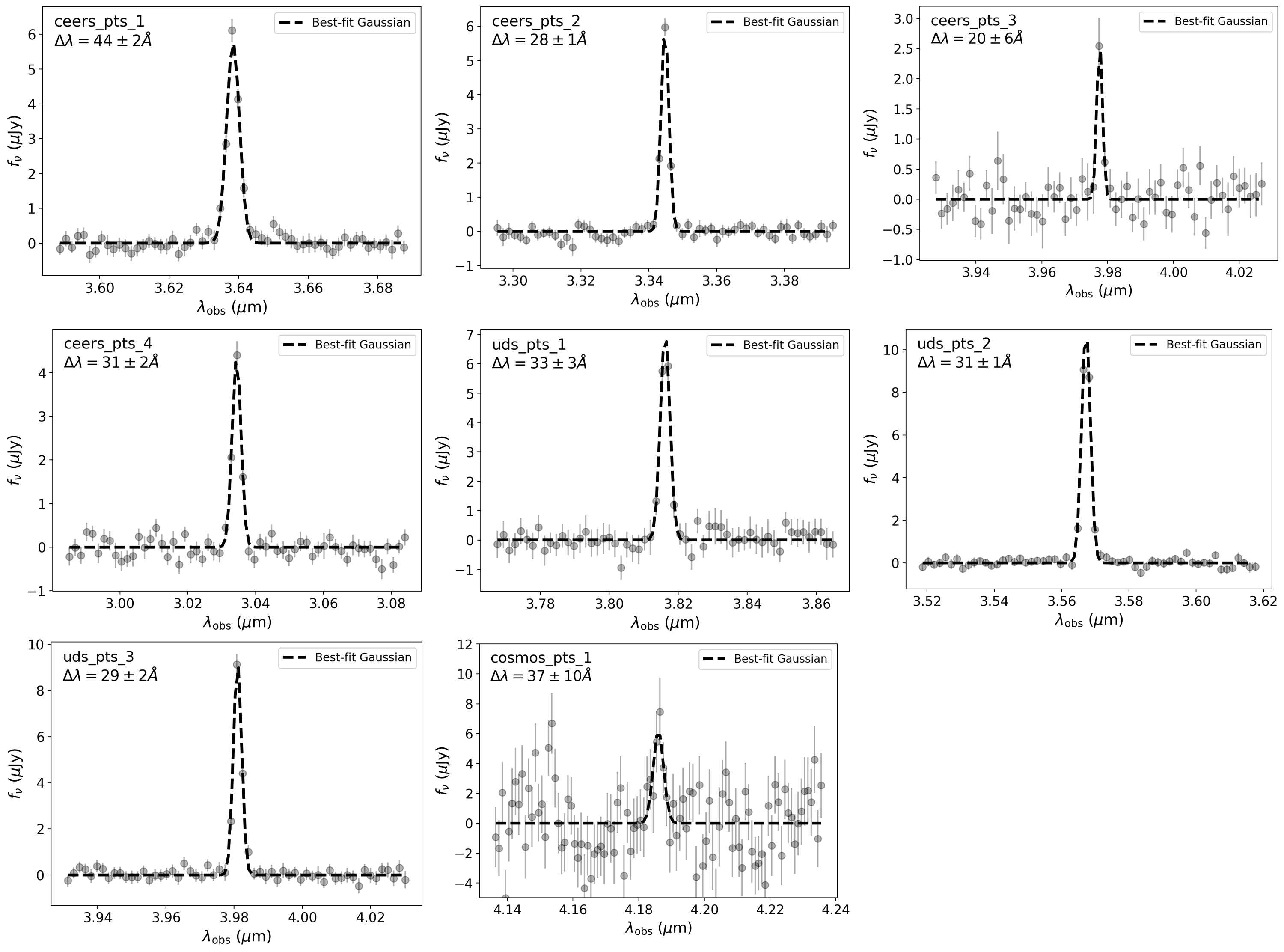}
    \raggedright
    \caption{Gaussian fits (dashed lines) to the H$\alpha$ line profiles
    (filled gray circles with error bars) in the medium-resolution spectra,
    which are the RUBIES Grating spectra for eight sources in the CEERS and 
    UDS fields and the COSMOS-3D NIRCam WFSS spectrum for the one in the
    COSMOS field. The FWHM widths of the best-fit Gaussian profile are quoted 
    as ``$\Delta\lambda$'' in \AA. }
    \label{fig:Halpha}
\end{figure*}

    The H$\alpha$ line widths were all measured using the medium-resolution 
spectra. Figure~\ref{fig:Halpha} shows the details of the Gaussian fitting to 
the H$\alpha$ line for each object. The high-S/N, $R\approx 1000$ Grating 
spectra of the objects in the CEERS and UDS fields allow us to conclude that 
their H$\alpha$ lines do not have a broad component. The NIRCam WFSS spectrum 
of the one in the COSMOS field does not have as high an S/N; nevertheless, its
H$\alpha$ line is still consistent with a single Gaussian profile of a narrow 
width. 

    The measured line widths are converted to velocities by
$\Delta v = c \Delta\lambda/\lambda_{obs}$, where $c$ is the speed of light,
$\Delta\lambda$ is the line FWHM, and $\lambda_{obs}$ is the line central
wavelength. These velocities are also reported in Table~\ref{tbl:lines}.

\begin{table*}[hbt!]
    \centering
    \tiny
    \caption{Emission line properties}
    \begin{tabular}{lccccccccccc}
        ID & & [O~\Romannum{2}]$\lambda$3727 & $\rm H\beta$ & [O~\Romannum{3}]$\lambda$4959 & [O~\Romannum{3}]$\lambda$5007 & $\rm H\alpha$ & [S~\Romannum{2}]$\lambda\lambda$6716,6731 & $c_{\rm slit}$ \\ \hline 
        ceers\_pts\_1 & $I$ & $1.78\pm0.32^\dagger$ & $2.21\pm0.41$ & $5.27\pm0.59$ & $16.27\pm0.91$ & $8.09\pm0.32$ & $0.42\pm0.19$ & $1.30$ \\ 
         & EW & ... & ... & $163.1\pm18.2$ & $513.0\pm28.7$ & $491.9\pm19.5$ & ... \\ 
         & $\Delta v$ & ... & ... & ... & ... & $360\pm17$ & ... \\ 
         & $L$ & $1.01\pm0.18$ & $1.25\pm0.23$ & $2.98\pm0.33$ & $9.20\pm0.51$ & $4.57\pm0.18$ & $0.24\pm0.11$ \\ 
        \hline 
        ceers\_pts\_2 & $I$ & $<0.69$ & $1.11\pm0.42$ & ... & $12.19\pm0.48^*$ & $3.83\pm0.31$ & $<0.31$ & $2.01$ \\ 
         & EW & ... & ... & ... & $1484.6\pm58.5^*$ & $801.5\pm64.9$ & ... \\ 
         & $\Delta v$ & ... & ... & ... & ... & $246\pm9$ & ... \\ 
         & $L$ & $<0.31$ & $0.49\pm0.19$ & ... & $5.39\pm0.21^*$ & $1.69\pm0.14$ & $<0.14$ \\ 
        \hline 
        ceers\_pts\_3 & $I$ & $<0.74$ & $0.84\pm0.16^\dagger$ & $1.22\pm1.09$ & $3.57\pm0.76$ & $1.72\pm0.41$ & $0.48\pm0.19^\dagger$ & $1.23$ \\ 
         & EW & ... & ... & $404.4\pm361.3$ & $1206.2\pm256.7$ & $624.2\pm148.8$ & ... \\ 
         & $\Delta v$ & ... & ... & $196\pm89$ & $351\pm36$ & $151\pm45$ & ... \\ 
         & $L$ & $<0.54$ & $0.61\pm0.12$ & $0.89\pm0.80$ & $2.60\pm0.55$ & $1.25\pm0.30$ & $0.35\pm0.14$ \\ 
        \hline 
        ceers\_pts\_4 & $I$ & $<0.89$ & $1.32\pm0.67$ & ... & $11.41\pm0.66^*$ & $4.91\pm0.40$ & $<0.47$ & $1.14$ \\ 
         & EW & ... & ... & ... & $1423.2\pm82.4^*$ & $1553.4\pm126.5$ & ... \\
         & $\Delta v$ & ... & ... & ... & ... & $307\pm18$ & ... \\ 
         & $L$ & $<0.29$ & $0.43\pm0.22$ & ... & $3.76\pm0.22^*$ & $1.62\pm0.13$ & $<0.15$ \\ 
        \hline 
        uds\_pts\_1 & $I$ & $<1.39$ & $2.37\pm1.19$ & $3.32\pm0.77$ & $11.12\pm1.54$ & $5.92\pm0.49$ & $<0.65$ & $1.31$ \\ 
         & EW & ... & ... & $586.6\pm136.0$ & $3605.0\pm499.2$ & $3297.3\pm272.9$ & ... \\
         & $\Delta v$ & ... & ... & $209\pm53$ & $286\pm24$ & $259\pm21$ & ... \\ 
         & $L$ & $<0.90$ & $1.54\pm0.77$ & $2.15\pm0.50$ & $7.21\pm1.00$ & $3.84\pm0.32$ & $<0.42$ \\ 
        \hline 
        uds\_pts\_2 & $I$ & $<0.92$ & $2.19\pm0.19^\dagger$ & $6.80\pm0.58$ & $17.86\pm0.89$ & $8.62\pm0.31$ & $<0.33$ & $1.31$ \\ 
         & EW & ... & ... & $541.4\pm46.2$ & $1656.9\pm82.6$ & $1374.0\pm49.5$ & ... \\
         & $\Delta v$ & ... & ... & ... & ... & $258\pm7$ & ... \\ 
         & $L$ & $<0.49$ & $1.17\pm0.10$ & $3.63\pm0.31$ & $9.54\pm0.48$ & $4.61\pm0.17$ & $<0.18$ \\ 
        \hline 
        uds\_pts\_3 & $I$ & $1.22\pm0.25^\dagger$ & $1.78\pm0.61$ & $4.22\pm1.62$ & $13.44\pm0.79$ & $6.21\pm0.38$ & $<0.33$ & $3.78$ \\ 
         & EW & ... & ... & $488.3\pm187.4$ & $1585.6\pm93.1$ & $1546.3\pm94.6$ & ... \\ 
         & $\Delta v$ & ... & $277\pm75$ & $259\pm28$ & $300\pm8$ & $246\pm9$ & ... \\ 
         & $L$ & $0.89\pm0.18$ & $1.30\pm0.45$ & $3.09\pm1.18$ & $9.83\pm0.58$ & $4.54\pm0.28$ & $<0.24$ \\ 
        \hline 
        cosmos\_pts\_1 & $I$ & $1.27\pm0.43$ & $1.53\pm0.18$ & $4.85\pm0.33$ & $10.97\pm0.30$ & $5.92\pm0.23$ & $<0.30$ & $1.74$ \\ 
         & EW & ... & ... & $604.1\pm41.1$ & $1393.1\pm38.1$ & $2009.3\pm78.1$ & ... \\
         & $\Delta v$ & ... & ... & ... & ... & $268\pm75$ & ... \\ 
         & $L$ & $1.07\pm0.36$ & $1.29\pm0.15$ & $4.08\pm0.28$ & $9.24\pm0.25$ & $4.98\pm0.19$ & $<0.25$ \\ 
        \hline 
    \end{tabular}
    \raggedright
    \tablecomments{The integrated line fluxes $I$ are in the units of 
    $10^{-18}$~erg~cm$^{-2}$~s$^{-1}$, and the values marked by ``$\dagger$''
    indicate that these were obtained by sum instead of profile fitting (see
    Section 3.2). For non-detections, the 2~$\sigma$ upper limits are quoted.
    The rest-frame equivalent widths EW
    are in the units of \AA. The line widths $\Delta v$ were calculated using 
    the FWHM of the lines and are in $\rm km~s^{-1}$. The line luminosities 
    $L$ are in the unit of $10^8 L_\odot$, where 
    $L_\odot=3.827\times10^{33}$~erg~s$^{-1}$. For the [O~\Romannum{3}] 
    doublets that are not resolved, the combined values are reported under
    [O~\Romannum{3}]$\lambda$5007 and are marked by ``$^{*}$''.
    All line intensities and luminosities quoted here are without applying the
    slit-loss corrections ($c_{\rm slit}$ in the last column) or the dust 
    extinction correction (see Table~\ref{tbl:properties}). 
    }
    \label{tbl:lines}
\end{table*}

\section{Discussion}

    A key question about these point-like, narrow-line sources is whether they are 
AGNs or star-forming (SF) galaxies. We discuss below these two scenarios and 
their implications.
    
\subsection{AGN or not}

    Quasars, which are type~1 AGNs, are point sources. Narrow permitted lines
are characteristics of type~2 AGNs. Therefore, a natural question is whether
our point-like objects could be type~2 quasars. The existing rest-frame optical
selection of type~2 quasars 
\citep[e.g.,][]{Zakamska2003, Alexandroff2013, Yuan2016}, however, do not
impose a morphological criterion; the main arguments for such narrow-line 
objects being quasars are their quasar-like absolute magnitudes
($M_B<-23$~mag), high-ionization lines (such as [Ne~\Romannum{5}]), and/or 
high [O~\Romannum{3}$\lambda$5007] luminosities 
($L_{\lambda 5007}>3\times 10^8 L_\odot$). In fact, a few  type~2 quasars have 
been detected their host galaxies by the HST/ACS without using a coronagraph 
to block the light of the nuclei \citep[][]{Zakamska2006}. In contrast, if our
objects are type~2 quasars, their hosts must be extremely compact dwarf 
galaxies.

    None of our objects has high-excitation lines such as [Ne~\Romannum{5}]
detected in their spectra. In terms of absolute magnitudes, none of them 
qualify as quasars ($M_B$ ranging from $-18.3$ to $-21.4$~mag). Interestingly, 
however, the five objects whose [O~\Romannum{3}] doublet can be resolved
all have $L_{\lambda 5007}>3\times 10^8 L_\odot$ after the slit-loss
corrections, which would meet the quasar criterion. 

\begin{figure}
    \centering
    \includegraphics[width=\linewidth]{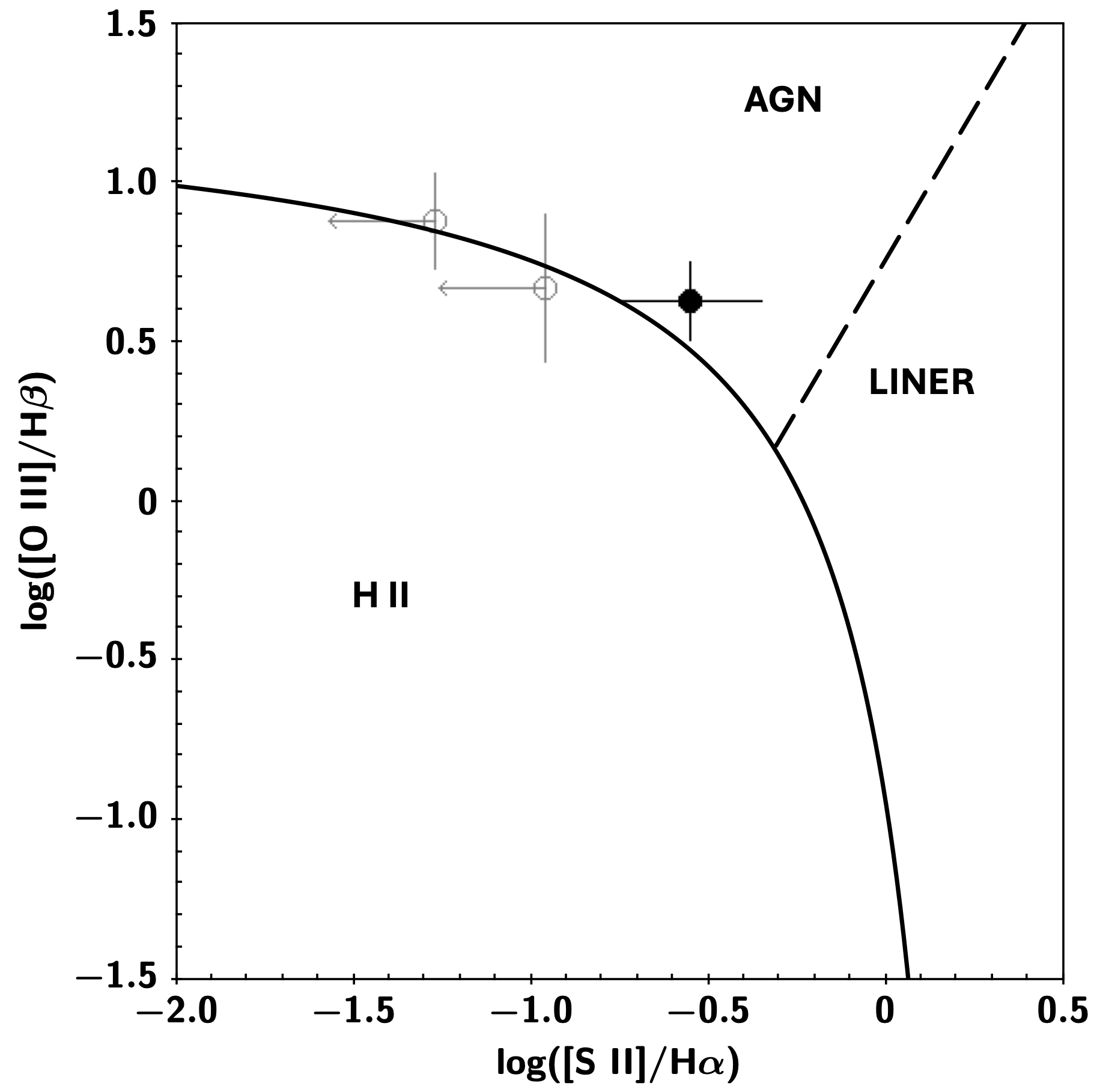}
    \caption{BPT diagram (following \citealt[][]{VO1987}) of the three objects 
that have lines covered by the existing data allowing such a diagnostics. 
The filled circle with error bars is \texttt{ceers\_pts\_3}. 
The two open circles with log([S~\Romannum{2}]/H$\alpha$) upper 
limits are \texttt{uds\_pts\_1} and \texttt{uds\_pts\_3} (from right to left).
The solid curve and the dashed line follow Eqns. 2 and 9 of 
\citet[][]{Kewley2006}, respectively; the former separates AGNs and 
star-forming galaxies (above and below the curve, respectively),
while the dashed line separates Seyferts and LINERs (left and right of the
line, respectively).
    }
    \label{fig:bpt}
\end{figure}

    The more important question is whether the strong lines are due to AGN
or star formation. The standard method for the diagnostics is to use the 
BPT diagram \citep[][]{BPT1981} and the like \citep[][]{VO1987}.
The existing data only allow us to use the 
log([O~\Romannum{3}]$\lambda$5007/H$\beta$) versus 
log([S~\Romannum{2}]$\lambda\lambda$6717,6731/H$\alpha$) diagnostics of 
\citet[][]{VO1987}, or the so-called ``[S~\Romannum{2}] BPT diagram'', which 
can be done for three objects. Among these three, only one 
(\texttt{ceers\_pts\_3}) has the [S~\Romannum{2}] doublet detected 
(and unresolved; see Table~\ref{tbl:lines}). For the other two objects 
(\texttt{uds\_pts\_1} and \texttt{uds\_pts\_3}), we can only use the 
2~$\sigma$ upper limits as the constraints. The results are shown in 
Figure~\ref{fig:bpt}, where the separations of three areas are based on the
local relations of \citet[][]{Kewley2006}. At $z>1$--2, it is well known 
that the AGN/SF-galaxy dividing curve in the ``[N~\Romannum{2}] BPT 
diagram'' shifts significantly to higher 
log([O~\Romannum{3}]$\lambda$5007/H$\beta$) and 
log([N~\Romannum{2}]$\lambda\lambda$6717,6731/H$\alpha$) values as compared 
to the local relation
\citep[e.g.,][]{Steidel2014, Shapley2015, Shapley2019, Shapley2025, Lam2026, Clarke2026}; 
however, the offset in the ``[S~\Romannum{2}] BPT diagram'' is not as
notable and is shifted to below the local AGN/SF-galaxy dividing curve
\citep[e.g.,][]{Shapley2019, Shapley2025}. Assuming that such trends 
persist at $z\sim 4$--5, \texttt{ceers\_pts\_3} falls in the AGN region. The 
other two are inconclusive, but are more likely to be consistent with SF 
galaxies.

   Two of our targets, \texttt{ceers\_pts\_3} and \texttt{ceers\_pts\_4},
have additional NIRCam data taken by PIDs 5398 (PI J. Kartaltepe) and 
2279 (PI R. Naidu), respectively. The former was observed in all seven CEERS
filters (plus F090W) on UT 2025 June 22 and the latter was observed in F444W 
on 2023 May 3-4, which were $\sim$913 and $\sim$133 days after the
CEERS observations, respectively. We reduced these data as well
in the same way as described in Section~2.1 to check for their variability.
None of them varied in between the two epochs.

\subsection{Scenario of star-forming galaxies}

   As the above analysis does not give a definite answer on whether our 
objects are AGNs, we consider the possibility that they could be SF
galaxies. 

\subsubsection{SED fitting}

\begin{figure*}[hbt!]
    \centering
    \includegraphics[width=0.95\textwidth]{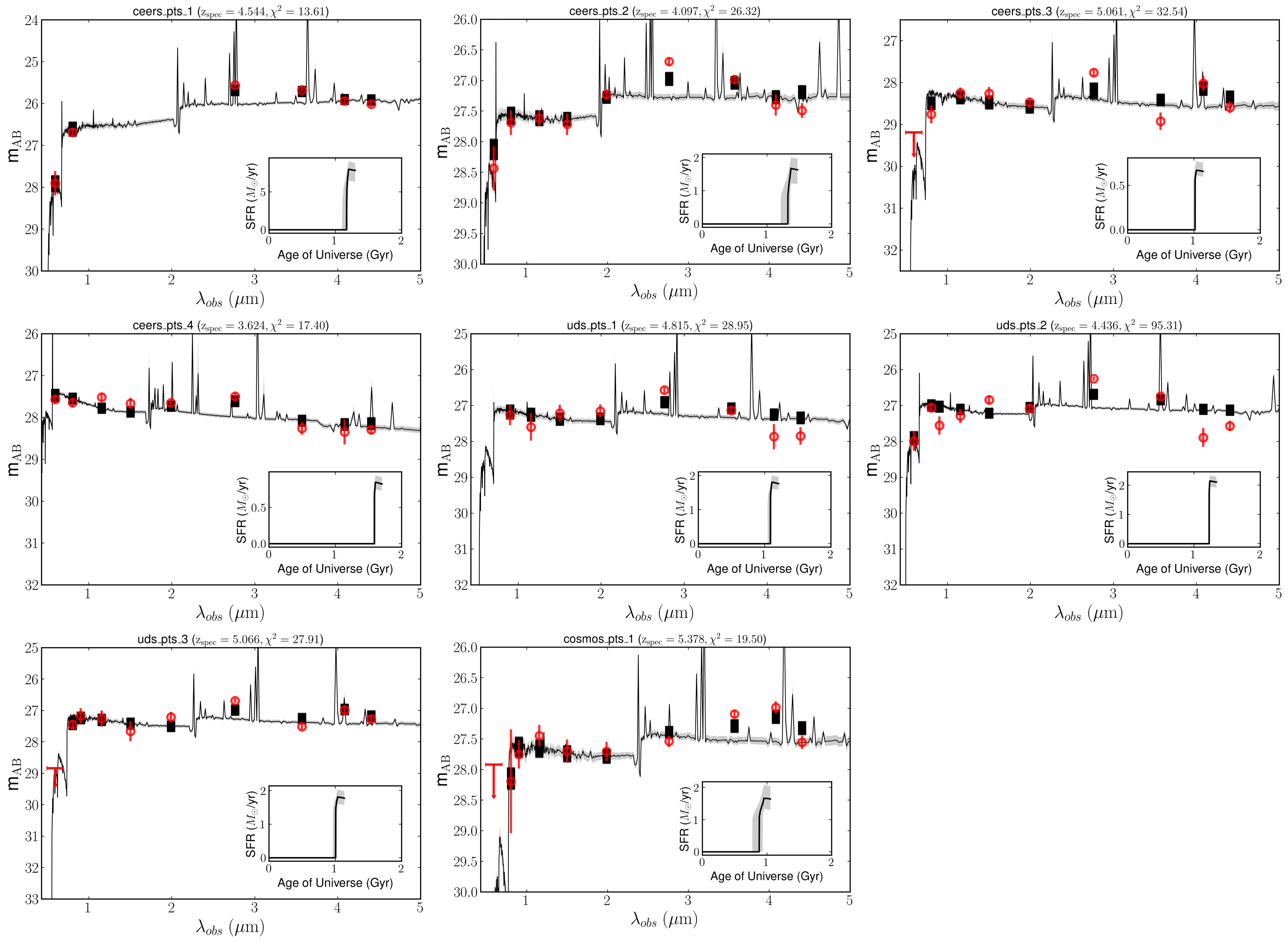}
    \raggedright
    \caption{{\sc Bagpipes} SED fitting of the eight objects, assuming that 
    they are SF galaxies. The redshifts are fixed at their $z_{\rm spec}$. The
    filled red circles with error bars are the SEDs. The posterior spectrum 
    corresponding to the median value is shown as the black curve in each
    panel, while the 16th to 84th percentile range is shown as the gray area 
    around it (almost cannot be discerned as they are narrow). The insets
    show the SFH.
    }
    \label{fig:sedfitting}
\end{figure*}

   Following \citet[][]{SY2025a}, the SEDs of our objects were constructed 
using the photometry on the 60mas images as described in Section~2.1. The
{\sc SExtractor} \texttt{MAG\_ISO} magnitudes were adopted (see 
Table~\ref{tbl:phot}).
These SEDs are fitted using the \textsc{Bagpipes} software 
\citep[][version 1.2.0]{Carnall2018}, which utilizes the stellar population 
synthesis models of \citet{Bruzual2003} with the initial mass function of
\citet{Kroupa2001}. The fitting is done at the targets' spectroscopic 
redshifts, and
we adopt an exponentially declining star formation history (SFH) where the star
formation rate SFR~$\propto e^{-t/\tau}$. The option to include nebular 
emission lines is enabled, and we use the Calzetti dust-extinction law 
\citep{Calzetti1994, Calzetti2001} with $A_{V}$ ranging from 0 to 8.0~mag.
The metallicity is allowed in the range of $0 \le Z_{*}/Z_{\odot} \le 2.5$, and
the ionization parameter can vary in $-4.0 \le {\rm log}(U) \le -2.0$.
The results of the SED fitting are summarized in Table~\ref{tbl:properties}, 
and Figure~\ref{fig:sedfitting} shows the 16th to 84th percentile range of 
the posterior spectra superposed on the SEDs. The insets show the evolution
of SFR. Interestingly, most of our objects indeed can be fitted reasonably as 
normal galaxies. The two exceptions are
\texttt{ceers\_pts\_3} (the one falling in the AGN region of the BPT diagram) 
and \texttt{uds\_pts\_2}.

    Assuming that these objects are galaxies, the most striking result from
the SED fitting is that they must all be new-born galaxies, with the ages of 
$\sim$110--170~Myrs. Most of them have stellar masses ($M_*$) on the order of 
$10^8M_\odot$, and only two are on the order of $10^9M_\odot$. They could have 
assembled their stellar masses secularly over their lifetimes at a modest SFR 
of a few $M_\odot$~yr$^{-1}$.

\begin{table*}[hbt!]
    \centering
    \small
    \caption{Galaxy properties inferred from SED fitting and emission line measurements}
    \begin{tabular}{l|ccccc|ccc}
    \hline\hline
        SID & $A_V$ & Age (Gyr) & $\log(M_*/M_\odot)$ & $Z/Z_\odot$ & SFR$_{\rm SED}$  & SFR$_{\rm H\alpha}$ & $E(B-V)_{\rm gas}$ & $f_{\rm ext}$ \\ \hline 
        ceers\_pts\_1 & $0.51_{-0.08}^{+0.07}$ & $0.13_{-0.02}^{+0.06}$ & $9.03_{-0.05}^{+0.09}$ & $0.24_{-0.06}^{+0.10}$ & $7.8_{-1.3}^{+1.1}$ & $12.2\pm0.5$ & $0.21\pm0.16$ & $1.33\pm0.29$ \\ 
        ceers\_pts\_2 & $0.30_{-0.13}^{+0.10}$ & $0.15_{-0.04}^{+0.11}$ & $8.41_{-0.08}^{+0.12}$ & $0.21_{-0.06}^{+0.09}$ & $1.6_{-0.5}^{+0.4}$ & $7.0\pm0.6$ & $0.16\pm0.33$ & $1.24\pm0.55$ \\ 
        ceers\_pts\_3 & $0.04_{-0.03}^{+0.05}$ & $0.12_{-0.01}^{+0.03}$ & $7.92_{-0.07}^{+0.11}$ & $0.53_{-0.19}^{+1.44}$ & $0.7_{-0.1}^{+0.1}$ & $3.2\pm0.8$ & ... & ... \\ 
        ceers\_pts\_4 & $0.08_{-0.05}^{+0.05}$ & $0.11_{-0.01}^{+0.02}$ & $7.99_{-0.05}^{+0.05}$ & $0.03_{-0.02}^{+0.02}$ & $0.8_{-0.1}^{+0.1}$ & $3.8\pm0.3$ & $0.22\pm0.44$ & $1.34\pm0.80$ \\ 
        \hline 
        uds\_pts\_1 & $0.05_{-0.04}^{+0.07}$ & $0.12_{-0.02}^{+0.04}$ & $8.35_{-0.06}^{+0.08}$ & $0.27_{-0.09}^{+0.13}$ & $1.8_{-0.2}^{+0.2}$ & $10.3\pm0.9$ & ... & ... \\ 
        uds\_pts\_2 & $0.13_{-0.05}^{+0.05}$ & $0.11_{-0.01}^{+0.02}$ & $8.38_{-0.04}^{+0.05}$ & $0.21_{-0.03}^{+0.05}$ & $2.1_{-0.2}^{+0.2}$ & $12.4\pm0.5$ & $0.27\pm0.08$ & $1.44\pm0.15$ \\ 
        uds\_pts\_3 & $0.05_{-0.04}^{+0.06}$ & $0.13_{-0.02}^{+0.04}$ & $8.39_{-0.07}^{+0.08}$ & $0.34_{-0.11}^{+0.16}$ & $1.8_{-0.2}^{+0.2}$ & $35.3\pm2.2$ & $0.17\pm0.30$ & $1.26\pm0.51$ \\
        \hline 
        cosmos\_pts\_1 & $0.15_{-0.09}^{+0.10}$ & $0.17_{-0.05}^{+0.12}$ & $8.47_{-0.10}^{+0.15}$ & $0.34_{-0.12}^{+0.21}$ & $1.6_{-0.3}^{+0.4}$ & $17.8\pm0.7$ & $0.26\pm0.10$ & $1.42\pm0.21$ \\ 
        \hline 
    \end{tabular}
    \raggedright
    \tablecomments{The stellar population properties inferred from the 
    {\sc Bagpipes} SED fitting and the gas-phase properties derived from the 
    line measurements are separated by a vertical line. ${\rm SFR_{SED}}$ and 
    ${\rm SFR_{H\alpha}}$ are the SFRs derived from the SED fitting
    (the instantaneous SFR from the last age bin, which has the width of
    5773 years) and the H$\alpha$ lines (see Equation 1), respectively, and 
    both are in units of $M_\odot$~yr$^{-1}$. The quoted ${\rm SFR_{H\alpha}}$ 
    values take into account the slit-loss corrections $c_{\rm slit}$ listed 
    in Table~\ref{tbl:lines} but not the dust extinction correction 
    $f_{\rm ext}$ as listed in the last column of this table.
    ${\rm E(B-V)_{gas}}$ is the gas-phase dust reddening derived using the
    Balmer decrement H$\alpha$/H$\beta$ and the Calzetti extinction law
    assuming the standard Case~B recombination. Two objects, 
    \texttt{ceers\_pts\_3} and \texttt{uds\_pts\_1}, cannot be derived their
    dust reddening in this way because they have H$\alpha$/H$\beta$ smaller
    than the canonical value of 2.86.
    }
    \label{tbl:properties}
\end{table*}

\subsubsection{H$\alpha$-based SFRs}

   The ``gold'' standard of deriving the instantaneous SFR of a star forming
galaxy is to use the H$\alpha$ line \citep[][]{Kennicutt1983, Kennicutt1998a}.
Here we follow the updated conversion of \citet[][]{KE2012}:
\begin{equation}
    {\rm SFR_{H\alpha}} =L({\rm H}\alpha)/1.862\times 10^{41},
\end{equation}
where $L(H\alpha)$ is the H$\alpha$ line luminosity in units of erg~s$^{-1}$.
We use the H$\alpha$ line intensities listed Table~\ref{tbl:lines} and
apply the slit-loss correction to calculate $L({\rm H}\alpha)$, and the derived
${\rm SFR_{H\alpha}}$ are listed in Table~\ref{tbl:properties}. These
are significantly higher than the SFRs derived based on the SED fitting 
(${\rm SFR_{SED}}$ in the table).

   Strictly speaking, both $L({\rm H}\alpha)$ and ${\rm SFR_{H\alpha}}$ should
be corrected for the dust extinction in gas. The common method is to use the
Balmer decrement H$\alpha$/H$\beta$ to estimate the extinction effect. We show 
below how this could be done for our objects, bearing in mind that the low S/N of 
the H$\beta$ lines leads to large errors in the extinction correction factors. 

   We adopt the Calzetti extinction law to calculate the gas-phase dust 
reddening following 
\begin{equation}\label{eq:ebv}
    E(B-V)_{\rm gas}=\frac{\log(R_{\rm obs}/R_{\rm int})}{0.4\cdot (k_{\rm H\beta}-k_{\rm H\alpha})},
\end{equation}
where $k_{\rm H\alpha}=3.327$ and $k_{\rm H\beta}=4.598$. $R_{\rm obs}$ is the 
observed line flux ratio between ${\rm H\alpha}$ and ${\rm H\beta}$, and 
$R_{\rm int}=2.86$ is the widely adopted intrinsic ratio in the Case B 
recombination at an electron temperature of $T_e=10^4$~K and an electron 
density of $n_e=10^2$~cm$^{-3}$ \citep[][]{Osterbrock1989}. The suppression of 
the line intensity due to extinction can then be expressed as 
\begin{equation}\label{eq:ext}
\begin{split}
    I(\lambda) & =I_0(\lambda) \cdot C_{\rm ext} \\
   & = I_0(\lambda)\cdot 10^{-0.4\cdot E(B-V)_{\rm gas} \cdot k_e(\lambda)},
\end{split}
\end{equation}
where $I_0(\lambda)$ and $I(\lambda)$ are the intensity before and after the
extinction by dust, respectively, and $k_e(\lambda)$ is calculated based on 
Eqns. 8a and 8b of \citet[][]{Calzetti2001}.

    A complication is the universality of the Case B assumption that was 
recently brought into question 
\citep[see][and the references there in]{SY2025b}. 
In fact, two of our objects, \texttt{ceers\_pts\_3} and
\texttt{uds\_pts\_1}, have $R_{\rm obs} < 2.86$. We ignore these two sources
and derive the dust extinction correction factors $f_{\rm ext}=1/C_{\rm ext}$ 
for the rest by still assuming the standard Case~B recombination. These factors
are listed in Table~\ref{tbl:properties}. With these corrections, the
H$\alpha$-based SFRs would be even higher. 

\subsubsection{Gas-phase metallicities}

    Three of our objects, namely, \texttt{ceers\_pts\_1}, 
\texttt{uds\_pts\_3}, and \texttt{cosmos\_pts\_1}, have [O~\Romannum{2}]
measurements, and therefore we can measure their metallicities using the
``R23'' method \citep[][]{Pagel1979}, where 
${\rm R23=([O~\Romannum{2}]+[O~\Romannum{3})/H\beta}$ and ${\rm logR23}$ can
be related to metallicity expressed in terms of 12+log(O/H). The correspondence
between ${\rm logR23}$ and 12+log(O/H), however, is not single-valued. Here we
adopt the calibration of \citet[][their Eqn. 7]{Jiang2019}. The dominant line 
in all these four objects is [O~\Romannum{3}$\lambda\lambda$4959,5007, which
is very close to H$\beta$ so that they suffer from nearly the same amount of
dust extinction. To avoid the large errors in the dust extinction correction
factors introduced by the low S/N H$\beta$ line, we choose not to apply this
correction here. For these four objects, we obtain 12+log(O/H) of 
$8.08\pm0.44$, 
$8.05\pm0.67$, and $8.29\pm 1.50$, respectively.
Comparing to the most commonly adopted solar oxygen abundance value of 
12+log(O/H)~$=8.69\pm0.03$ \citep[][]{Amarsi2018} or $8.75\pm0.03$
\citep[][]{Bergemann2021}, our objects have significantly lower oxygen
abundances, which is broadly consistent with their young ages. 
    
\subsection{Compare to ``green pea'' galaxies at low redshifts}

    The closest resemblances to our objects are probably the so-called
``green pea'' galaxies \citep[GPs;][]{Cardamone2009}. GPs are compact
galaxies at low redshifts ($z\lesssim 0.4$) that have extremely strong
[O~\Romannum{3}] emission lines, which make them green in the color
composite of optical images.

    Green peas are a heterogeneous group and have both AGNs and SF galaxies, 
with the latter being the majority. The SF GPs have ${\rm SFR} \sim 10 M_\odot$~yr$^{-1}$ and low reddening of ${\rm E(B-V)\lesssim 0.25}$, which are 
quite similar to our objects; they have stellar masses 
$M^* \approx 10^{8.5-10}M_\odot$ and metallicities 12+log(O/H)~$\sim 8.7$, 
both of which are significantly higher than our objects but are still close 
\citep[e.g.,][]{Cardamone2009}.

    However, there are still notable differences between our objects and the
general GP population besides their redshifts. The most obvious one is that our 
objects are point-like but most GPs are not. While GPs appear to be unresolved in 
ground-based images, almost all that have HST images show that their GP regions 
are embedded in extended hosts, which often have complex morphologies indicative 
of mergers
\citep[e.g.,][]{Cardamone2009, Amorin2012, Izotov2018, Kim2020, Clarke2021, Keel2022, Flury2022a}.
Surface brightness dimming is not likely to explain the lack of extended emissions 
around our objects if they have similar features as those of most GPs. This is
demonstrated by the simulations presented in Appendix~\ref{sec:gpsimu}, where the
HST/ACS images of the low-redshift GPs from \citet{Keel2022} are artificially 
shifted to $z=4$ and applied surface brightness dimming. Unless a GP has a 
point-like core in the first place, its simulated image at $z=4$ would not be 
point-like. Among the 33 GPs of \citet{Keel2022} observed with HST/ACS, only two
of them have such point-like cores. In this sense, only a very small minority of
GPs could be regarded as analogs of our objects.

    Another major difference is in their ages. The investigations of the GP 
host galaxies all conclude that they are several Gyrs old and that the GP 
regions are SF regions newly formed from within \citep[e.g.,][]{Amorin2012, Clarke2021, Paswan2022}. The SF regions themselves are likely also a mixture
of young and old stars. This is best shown in the study of the GP galaxy at $z=0.0472$ of \citet[][]{Paswan2022}
\footnote{This is a so-called ``blueberry'' galaxy, a nickname used to refer
to the GPs in the local universe \citep[][]{Yang2017}.}, 
where the GP's SF region even has Mg~\Romannum{1}~$\lambda$5173 absorption line 
characteristics of late type
stars; based on this feature, those authors derived an average mass-weighted age 
of $\sim$5~Gyrs in this region. In contrast, our objects, if they are
SF galaxies, all have young ages of $\sim$110-170~Myrs (see
Table~\ref{tbl:properties}), i.e., they are newly formed. 

\section{Summary and Conclusion}

    In this work, we report the discovery of a potentially new population of 
point-like, narrow-emission line objects using the public JWST imaging and 
spectroscopic data in three wide survey fields. Our sample consists of eight 
objects at $z=3.624$--5.061, which all have narrow H$\alpha$ line with widths of 
only 150--360~km~s$^{-1}$. Their light profiles can be well described by 2D 
Gaussian functions, and the FWHM values are only 3.7\%--34.6\% larger than those 
of the PSFs. In the bluest bands where the spatial resolutions are the best, 
their FWHM sizes correspond to 0.49 to 0.96~kpc (the effective radii $R_e$ being 
half of these values). While their possessing strong emission lines 
(in particular the [O~\Romannum{3}] lines) resembles the ``green pea'' galaxies
at $z\lesssim 0.4$, our objects are different in that they are not embedded in
extended host galaxies made of old stellar populations; and their lack of
extended emissions is likely genuine and cannot be explained by the surface
brightness dimming.

    Due to the limitation of the existing data, the exact nature of these 
objects is still unclear. Their point-like morphologies and narrow H$\alpha$ 
lines naturally lead to the speculation whether they could be type~2 quasars. 
There are arguments either supporting or disfavoring this scenario, however.
While their continuum luminosities are too low ($M_B\approx -18.3$ to 
$-21.4$~mag) to qualify as quasars, they all have strong [O~\Romannum{3}] 
lines; for the six objects whose [O~\Romannum{3}]$\lambda$5007 line can be 
resolved, the luminosities of this line are all above $3\times 10^8 L_\odot$ 
and in the quasar regime. One of these six has the [S~\Romannum{2}] line 
detected, and two others have upper limits of this line. The BPT-diagram
diagnostics of these three shows that the one with [S~\Romannum{2}] detection 
is likely AGN but the other two are inconclusive due to their 
[S~\Romannum{2}] measurements only being upper limits as well as the large 
errors in the measurements of the H$\beta$ line.
Two objects among the eight in our sample (one being the object with 
[S~\Romannum{2}] detection) have additional NIRCam data taken at later times
to check for variability, and none of them varied.

    On the other hand, most of these eight objects can be fitted reasonably
with normal galaxy templates and are consistent with being very young SF 
galaxies with ages of $\sim$110--170~Myrs (median 120~Myrs). The stellar 
masses and the instantaneous SFRs inferred from the SED fitting range from 
$10^{7.9-9.0}M_\odot$ (median of $10^{8.4}M_\odot$) and 
0.7--13.7~$M_\odot$~yr$^{-1}$ (median of 1.8~$M_\odot$~yr$^{-1}$), 
respectively. The instantaneous SFRs derived from 
their H$\alpha$ lines before the dust extinction correction, however, range 
from 3.2 to 50.1~$M_\odot$~yr$^{-1}$ (median of 12.2~$M_\odot$~yr$^{-1}$). 

    Regardless of their exact nature, this population of point-like, 
narrow-line objects deserve further investigations, and deeper, 
medium-resolution spectroscopy will be critical in the future diagnostics. If 
they are confirmed to be AGNs, they will constitute a new kind of type~2 AGNs 
that are low-luminosity and almost ``hostless''. If they are confirmed to be SF
galaxies, they likewise will constitute a new kind because of their point-like 
morphologies and young ages, which imply that they began their star 
formation secularly (and in isolation) from a very compact core.

\begin{acknowledgements}

We thank William Keel for kindly offering the reduced ACS images of the GP
galaxies used for comparison in this work.
We acknowledge the support from the University of Missouri Research 
Council grant URC-23-029 and
National Science Foundation Grant No.\ AST-2307447.

{The JWST and HST data presented in this article were obtained from the 
Mikulski Archive for Space Telescopes (MAST) 
at the Space Telescope Science Institute. 
The specific observations analyzed can be accessed via 
\dataset[doi:10.17909/26re-7589]{http://dx.doi.org/10.17909/a1ra-cq50}.
}
 
\end{acknowledgements}

\appendix
\counterwithin{figure}{section}
\counterwithin{table}{section}

\section{NIRCam Empirical PSFs}\label{sec:epsf}

   To best assess the sizes of the NIRCam PSFs, we constructed empirical PSFs 
in the relevant NIRCam bands. We carefully selected 20 isolated stars at 
$\sim$23.5--24.5~mag in each field, which are at the faintest brightness level
of sufficient S/N so that their point-source nature (showing diffraction spikes) 
is clear by visual inspection. However, the empirical PSFs were not made by simply 
stacking their images on the 60mas mosaics of the entire wide field, because these 
PSF stars could be artificially distorted on these large images if their locations 
are far away from the tangential point of the projection. We took a more 
appropriate approach.

    For each star, we combined the individual exposures (the ``cal.fits''
files) containing this star using the {\tt ResampleStep} in the 
{\tt calwebb\_image3} module of the JWST pipeline, setting the tangential 
point of the projection to the position of the star.
The pixel scale of the SW bands (F090W, F115W, F150W, and F200W) was set
to 20 mas (i.e., a factor of $\sim$1.55 subsampling of the SW detector
native pixel size), while that of the LW bands (F277W, F356W, F410M, and
F444W) was set to 30 mas (i.e., a factor of $\sim$2.10 subsampling of the
LW detector native pixel size). Based on the NIRCam performance as reported
at the JWST User Documentation site (JDoc)
\footnote{\url{https://jwst-docs.stsci.edu/jwst-near-infrared-camera/nircam-performance/nircam-point-spread-functions}}, 
these choices were to subsample the PSFs of F200W and F277W by 
a factor of $\sim$3, which are the two bands that our targets most often have 
sufficient S/N for the light profile measurements (see Section~3.1).

    From these images, we made a set of 101$\times$101-pixel 
cutouts around the PSF stars and masked all other sources. These star 
images were then combined using the \texttt{EPSFBuilder} routine of 
{\sc photutils} \citep[][v2.2.0]{larry_bradley_2023} to create the empirical PSF 
in a given band in each field. 

\begin{figure*}
    \centering
    \includegraphics[width=0.95\textwidth,height=0.95\textheight,keepaspectratio]{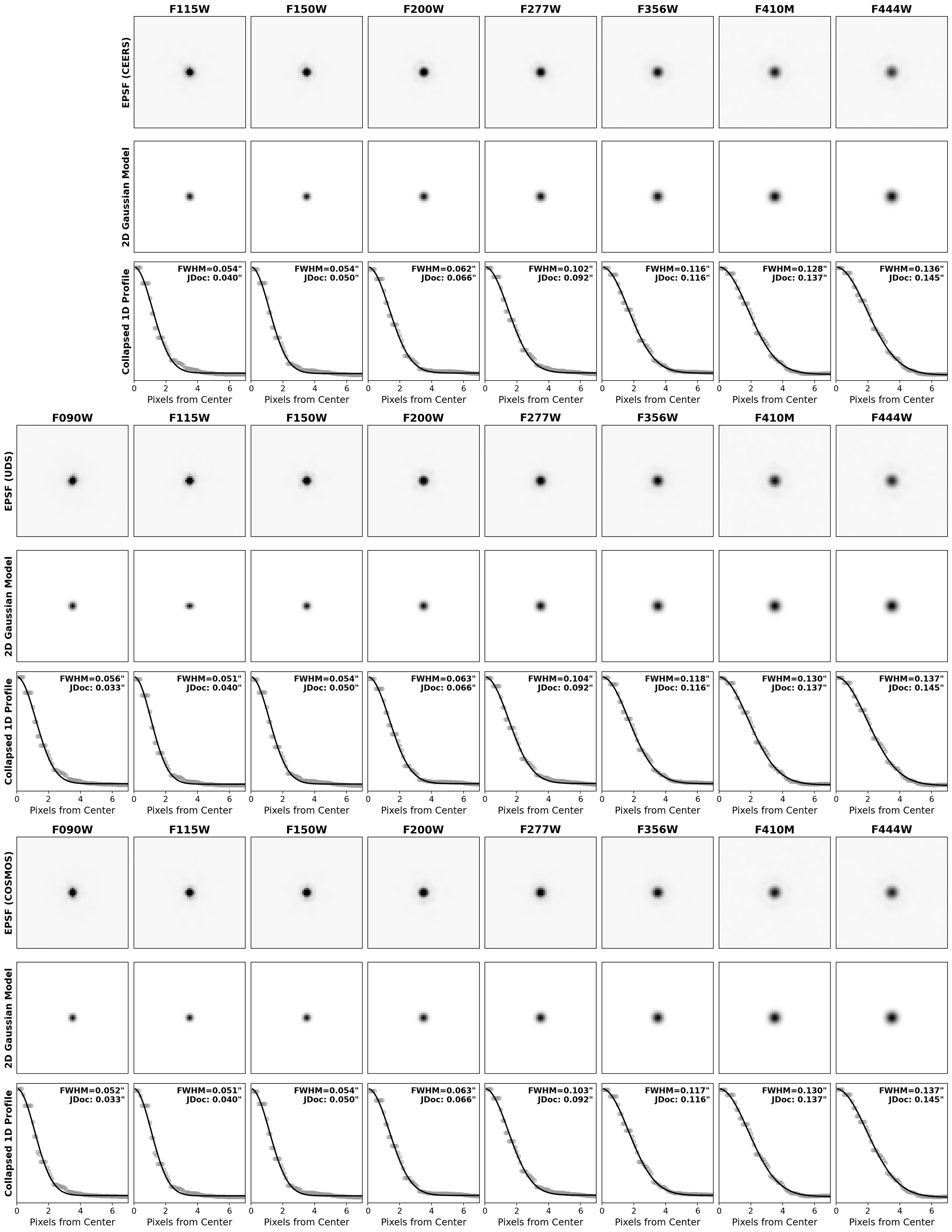}
    \raggedright
    \caption{Empirical PSFs in the NIRam bands as labeled. From top to bottom,
    the results are shown for the CEERS, UDS and COSMOS fields. For each field,
    the first row shows the EPSFs that we derived, the second row shows the
    2D Gaussian model images, and the third row shows the best Gaussian fits (black
    curves) to the light profile (gray circles). The measured FWHM values are 
    compared to those provided at the JDoc site.
    }
\label{fig:epsf}
\end{figure*}

    To quantify the PSF sizes, we use a 2D Gaussian model to fit these 
empirical PSFs, and the results are shown in Figure~\ref{fig:epsf}. The 
measured FWHM values are given in Table~\ref{tbl:epsf}, together with the 
values reported in JDoc for comparison. Our measurements are rather consistent 
in the three fields, and there are some notable differences as compared to the 
JDoc values. For the purpose of this work, we adopt our PSF size measurements 
in each field when comparing to the target sizes. 

\begin{table}[]
    \centering
    \caption{Comparison of the empirical PSF FWHM sizes and the JDoc values}
    \begin{tabular}{ccccccccc} \hline 
         Filter & F090W & F115W & F150W & F200W & F277W & F356W & F410M & F444W \\ \hline 
         CEERS & ... & 0\arcsec.054 & 0\arcsec.054 & 0\arcsec.062 & 0\arcsec.102 & 0\arcsec.116 & 0\arcsec.128 & 0\arcsec.136 \\
         UDS & 0\arcsec.056 & 0\arcsec.051 & 0\arcsec.054 & 0\arcsec.063 & 0\arcsec.104 & 0\arcsec.118 & 0\arcsec.130 & 0\arcsec.137 \\ 
         COSMOS & 0\arcsec.052 & 0\arcsec.051 & 0\arcsec.054 & 0\arcsec.063 & 0\arcsec.103 & 0\arcsec.117 & 0\arcsec.130 & 0\arcsec.137 \\
         JDoc & 0\arcsec.033 & 0\arcsec.040 & 0\arcsec.050 & 0\arcsec.066 & 0\arcsec.092 & 0\arcsec.116 & 0\arcsec.137 & 0\arcsec.145\\ 
         \hline 
    \end{tabular}
    \label{tbl:epsf}
\end{table}

\section{Geometric Slit-loss Corrections for Point-like sources}
\label{sec:slitloss}

    Deciding the NIRSpec MSA slit placements needs to optimize among a series
of targets. As the MSA shutter positions are not adjustable, it often happens
that a slit formed by the open shutters cannot fully cover a desired target 
even if the target is small as compared to the slit width. Therefore, the 
line intensities measured from the spectrum obtained in such a case are only a
fraction of the true line intensities. To correct this kind of geometric slit
losses for our point-like sources, our approach is as follows. 

\begin{figure}
    \centering
    \includegraphics[width=0.5\linewidth]{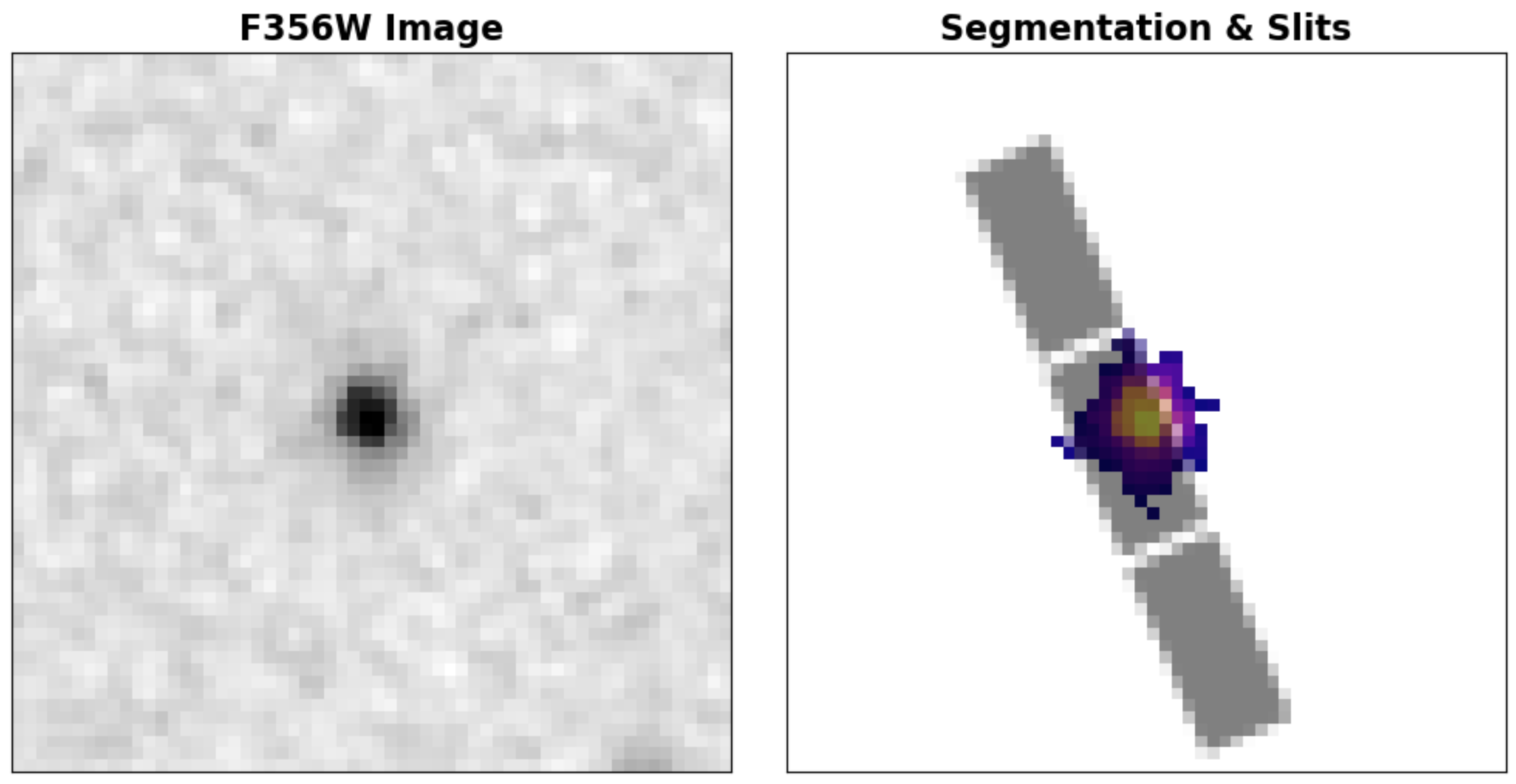}
    \caption{Demonstration of slit-loss correction factor estimation. 
    Left: F356W image stamp centered on the target (\texttt{uds\_pts\_2}). 
    Right: segmentation map of the source with the footprints of the three open 
    MSA shutters (forming the slit for this target) overlaid in gray. 
    }
    \label{fig:slit-corr}
\end{figure}

   We used the F356W images, which cover the H$\alpha$ lines at the redshifts of 
our targets, as the basis to estimate the correction factors. We generated a 
segmentation map for each object using {\sc photutils}, adopting a source 
detection threshold of 2.0
times the background root mean square values. The total source flux $f_{\rm src}$ 
was measured by summing the background-subtracted values within this segmentation 
map. We then generated a slit mask using the slit coverage (gray rectangles in 
Figure~\ref{fig:slit-corr}), and the flux falling inside the MSA slits 
$f_{\rm slit}$ was then computed as the sum of source pixels covered by the slit 
mask. We adopted the ratio of $c_{\rm slit}=f_{\rm src}/f_{\rm slit}$ as the 
multiplicative slit-loss correction factor, which are reported in the last column 
of Table~\ref{tbl:lines}. 

\section{Simulation of Green Pea Galaxies}
\label{sec:gpsimu}

    Could low-redshift GP galaxies be point-like due to surface brightness
dimming if they were at redshifts comparable to those of our targets? To answer 
this question, we carried out simulations using the GP galaxies in 
\citet[][]{Keel2022}, which have ACS images in F775W or F850LP from the HST Cycle 
30 program ID 15445 (PI. W. Keel). The reduced images have the pixel scale of
0\farcs05~pix$^{-1}$. We used the {\sc FERENGI} code
\footnote{The code was written in the Interactive Data Language (IDL), and we
executed it within the Python environment using the Python-to-IDL bridge.}
of \citet[][]{Barden2008} to redshift these ACS images to $z=4$ and in the NIRCam 
F444W band. The empirical ACS PSFs were derived using the point sources in the ACS 
images. 

\begin{figure}
    \centering
    \includegraphics[width=\linewidth]{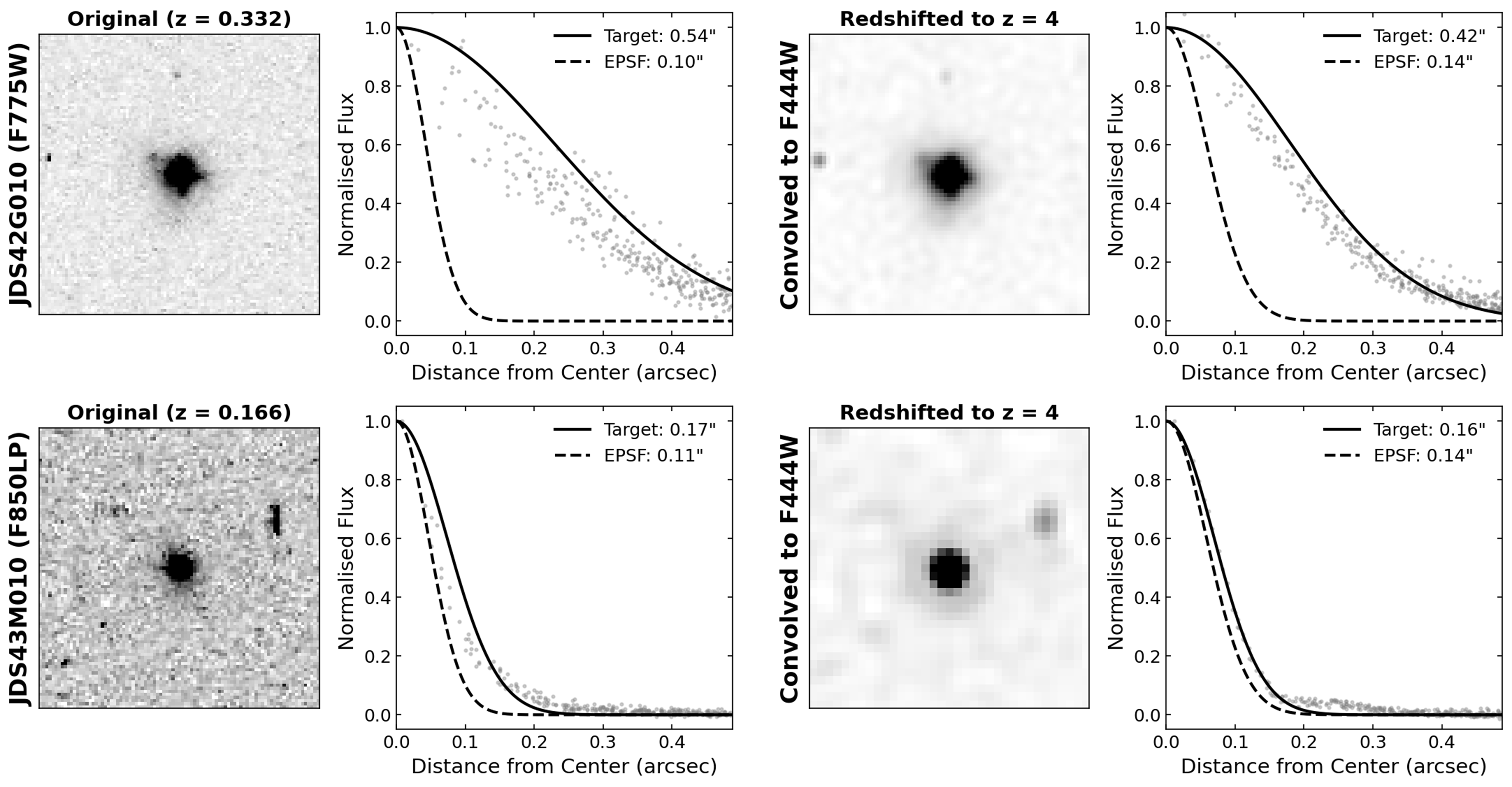}
    \caption{Two examples of GP galaxy surface brightness dimming simulations, one
in each row. The object ID, redshift and ACS band (from \citealt[][]{Keel2022}) are 
as labeled. The top row is for a compact GP galaxy that is well resolved and has an 
asymmetric light profile the, while the bottom row is for one of the only two 
point-like GP galaxies in \citet[][]{Keel2022} (out of 33 total).
In each row, the first panel shows the object's original ACS image, 
while the second panel shows its light profile (grey points) and the best-fit 
Gaussian profile (solid curve). For comparison, the light profile of the empirical 
PSF is superposed (dashed curve). The FWHM values of the best-fit and the EPSF are 
labeled. The third and forth panels are similar to the previous two, but are for 
its simulated image at $z=4$ in NIRCam F444W and the corresponding light profile.
}
    \label{fig:gbsimu}
\end{figure}

   As it turns out, none of those that have complex morphologies are point-like in 
the simulated images. These include the objects that are compact but still have 
asymmetric light profiles, one of which is shown in the top panel of
Figure~\ref{fig:gbsimu} for demonstration. Among the 33 GP galaxies, only two are 
point-like in the simulated images, and \emph{they are both point-like in the 
original ACS images} (objects ``JDS43Q'' and ``JDS43M''). One of them is shown in 
the bottom panel of Figure~\ref{fig:gbsimu}.

    In short, surface brightness dimming would not convert a GP galaxy of complex
morphology into a point-like source. Based on the Keel et al's sample, only a very 
small fraction of GP galaxies (2 out of 33, or $\sim$6\%) are point-like and could
be the low-redshift analogs of our objects.

\end{document}